\documentclass[a4paper]{article}
\usepackage{ASVspoof}
\usepackage{epsfig,amssymb,amsmath}
\usepackage{hyperref}
\usepackage{algorithm}
\usepackage{algorithmic}
\usepackage{amsmath}
\usepackage{pifont}
\usepackage{tikz}
\usepackage{cite}
\ninept

\title{Temporal Variability and Multi-Viewed Self-Supervised Representations to Tackle the ASVspoof5 Deepfake Challenge}

\makeatletter
\def\name#1{\gdef\@name{#1\\}}
\makeatother
\name{{\em Yuankun Xie$^1$\textsuperscript{,}$^2$, Xiaopeng Wang$^2$\textsuperscript{,}$^3$, Zhiyong Wang$^2$\textsuperscript{,}$^3$, Ruibo Fu$^2$, Zhengqi Wen$^2$,}\\
	      {\em Haonan Cheng$^1$, Long Ye$^1$}}
\address{
$^1$ State Key Laboratory of Media Convergence and Communication, \\Communication University of China\\
$^2$ Institute of Automation, Chinese Academy of Sciences\\
$^3$ School of Artificial Intelligence, University of Chinese Academy of Sciences\\ 
{\small \tt xieyuankun@cuc.edu.cn}}

\begin{document}
\maketitle

\begin{abstract}

ASVspoof5, the fifth edition of the ASVspoof series, is one of the largest global audio security challenges. It aims to advance the development of countermeasure (CM) to discriminate bonafide and spoofed speech utterances. In this paper, we focus on addressing the problem of open-domain audio deepfake detection, which corresponds directly to the ASVspoof5 Track1 open condition. At first, we comprehensively investigate various CM on ASVspoof5, including data expansion, data augmentation, and self-supervised learning (SSL) features. Due to the high-frequency gaps characteristic of the ASVspoof5 dataset, we introduce Frequency Mask, a data augmentation method that masks specific frequency bands to improve CM robustness. Combining various scale of temporal information with multiple SSL features, our experiments achieved a minDCF of 0.0158 and an EER of 0.55\% on the ASVspoof 5 Track 1 evaluation progress set.

\end{abstract}

\section{Introduction}
With the rapid advancement of text-to-speech (TTS) and voice conversion (VC) technologies, deepfake speech has proliferated significantly, making it increasingly difficult for humans to discern. The latest generation models \cite{chen2024vall}, benefiting from audio language models, have even achieved human parity in zero-shot TTS synthesis for the first time. This poses significant threats, including fraud, misleading public opinion, and privacy violations. Therefore, the urgent development of effective countermeasure (CM) technologies is crucial.

The automatic speaker verification spoofing and countermeasures (ASVspoof) challenge has been at the forefront of developing speech spoofing detection tasks. The common goal is to foster progress in the development of CMs, which are capable of discriminating between real and spoofed speech utterances. ASVspoof has successfully held four editions since 2015 \cite{wu15e_interspeech,kinnunen17_interspeech,todisco19_interspeech,yamagishi21_asvspoof} and has garnered significant attention. Especially around ASVspoof2019 and ASVspoof2021, many outstanding works have been proposed to address the problem of audio deepfake detection (ADD) \cite{lee22q_interspeech,xie23c_interspeech,wang2024generalized,wang2024genuine}.
Recently, ASVspoof5 was focuses on deepfake detection, dividing into two tracks: the deepfake detection track, which is independent of ASV tasks, and the SASV task, which is related to ASV tasks. 

To develop CMs for ADD, we focus on Track 1, a stand-alone speech deepfake (bona fide vs spoof) detection task. For Track 1 of ASVspoof5, there are two conditions: closed and open. The closed condition requires not using any data outside of the ASVspoof5 dataset or any pre-trained models. The open condition allows the use of additional datasets and pre-trained features, but these data resource cannot overlap with the source domains of ASVspoof5, meaning they cannot include data from the Multilingual Librispeech (MLS), LibriLight, or MUSAN speech subset.
We believe the design of the closed and open condition is excellent, as it prohibits overlap between training and test data, including pre-trained features, backbone networks, and data. This is a novel concept in the field of domain generalization known as test data information leakage \cite{yu2024rethinking}. Avoiding leakage is crucial to effectively measure the generalization capability of the CM.

In this paper, we conduct a series of studies on track 1 under open condition. Specifically, we investigate ASVspoof5 from various aspects including data expansion, data augmentation (DA) and self-supervised learning (SSL) feature extraction. For data expansion, we first explored the effectiveness of additional data, including the classic ASVspoof previous series, the latest audio deepfake detection (ADD) datasets such as MLAAD \cite{muller2024mlaad}, and Codecfake \cite{xie2024codecfake}. As data augmentation, we investigated the effectiveness of traditional front-end DA methods, including low-pass filtering, high-pass filtering, time and pitch duration changes, MUSAN \cite{David2015MUSAN} and RIR \cite{Tom2017A} noise augmentation. Considering the high-frequency bands gap characteristic of the ASVspoof5, we propose a novel augmentation method called Frequency mask (Freqmask), which randomly masks frequency bands to enhance the robustness of CM. For SSL feature extraction, we exhaustively investigated the performance of the available pre-trained self-supervised features on ASVspoof5 and explored their performance under different fixed and variable lengths. After conducting the aforementioned studies, we integrated seven CM methods at different scales through logits score fusion, considering both temporal information and diverse SSL categorical perspectives. Ultimately, we achieved a minimum detection cost function (minDCF) of 0.0158 and an equal error rate (EER) of 0.55\% on the ASVspoof5 evaluation progress set.

\section{Countermeasure}
In this section, we will sequentially introduce the approaches we considered, covering data expansion, data augmentation, SSL feature selection, and backbone network, from data pre-processing to training.

\subsection{Data expansion}
In the open condition, expanding the training data is considered permissible, making additional data expansion a key consideration to enhance the generalization of the CM. We consider to select several representative datasets to co-train with the ASVspoof5 training set. Our selection criteria ensure that the datasets do not overlap with the ASVspoof5 source domain (matching the organizers' rules), are relatively clean (comparable to the ASVspoof5 test set), and have representative spoofing techniques (to enhance CM generalization). 

\textbf{ASVspoof2019LA (19LA)}: The 19LA training and development sets include six spoofing methods (A01-A06), while the test set includes thirteen spoofing methods (A07-A19). To ensure diversity in the spoofing techniques of the training data, we selected both the training and test sets of 19LA as the augmentation datasets, comprising a total of 96,617 speech samples with 19 spoofing methods.

\textbf{MLAAD}: The MLAAD dataset currently includes the most spoofing techniques, utilizing 54 TTS models and comprising 21 different architectures, with a total of 76,000 speech samples.

\textbf{Codecfake}: The codecfake dataset is designed for Audio Language Model (ALM) based audio detection, containing 1,058,216 audio samples synthesized using seven different neural codecs. Given its significantly larger size compared to the ASVspoof5 training set, we selected a subset of 79,369 audio samples from the test set (C1-C6) for the co-training experiment.
\begin{figure}[t]
	\centering
	\includegraphics[width=3.1in]{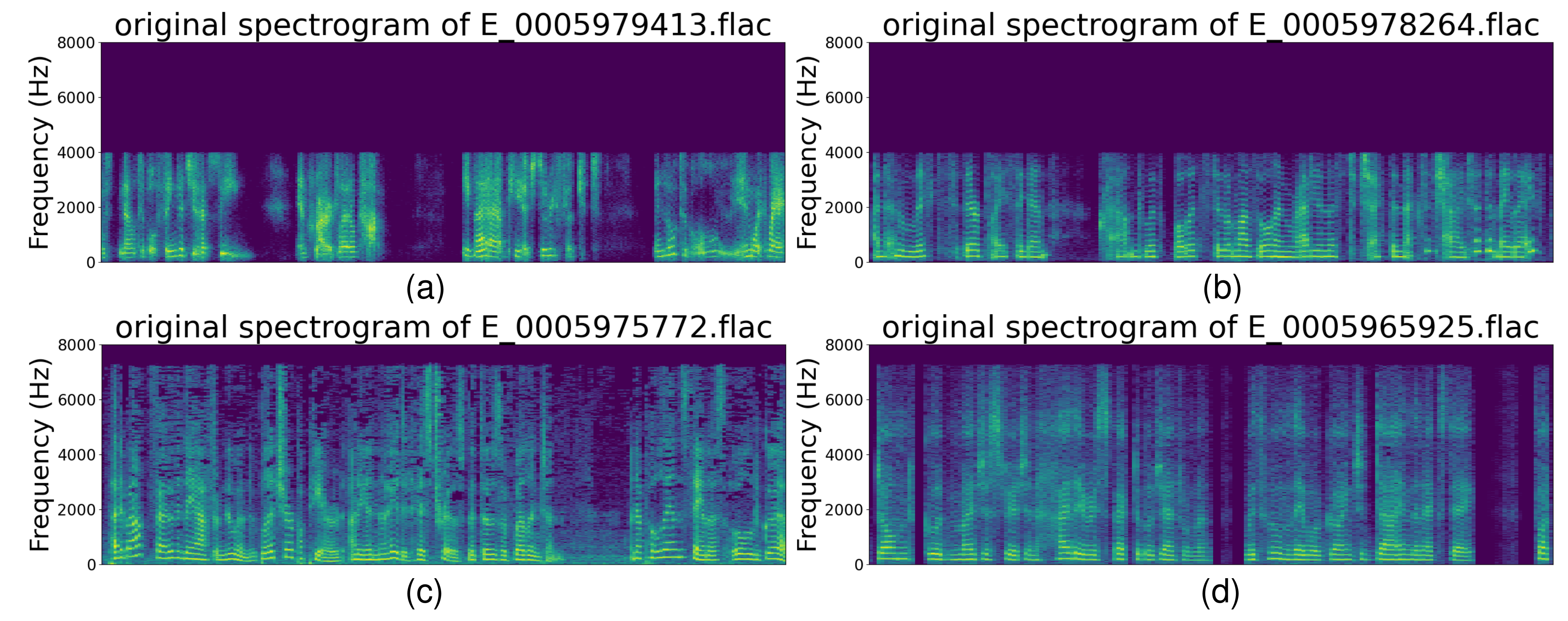}
	\caption{The original spectrogram of ASVspoof5 evaluation set shows high-frequency band gaps. (a) and (b) exhibit gaps above 4kHz, while (c) and (d) exhibit gaps above 7kHz.}
	\label{fig:orispec}
\end{figure}

\begin{figure}[t]
	\centering
	\includegraphics[width=3.1in]{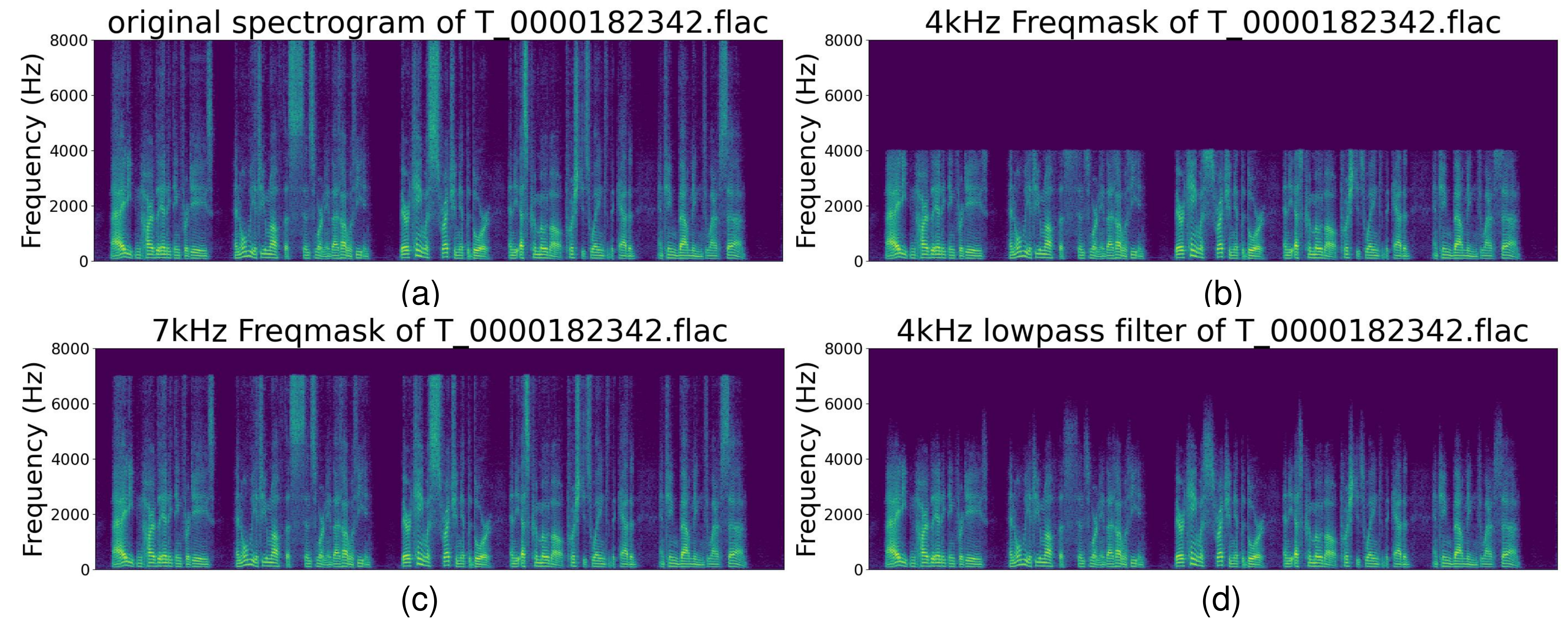}
	\caption{Comparison between the DA techniques using Freqmask and a low-pass filter. (a) shows the original spectrogram, (b) and (c) show the spectrograms with 4 kHz and 7 kHz Freqmask applied, respectively, and (d) shows the spectrogram with a 4 kHz low-pass filter applied.}
	\label{fig:freqmask}
\end{figure}
\subsection{Data augmentation}
\subsubsection{Freqmask data augmentation method}
As shown in Figure \ref{fig:orispec}, a random selection of samples from the ASVspoof5 evaluation set reveals high frequency bands missing. We refer to this phenomenon as high-frequency band gaps. This frequency band gap phenomenon aggressively removes values in the higher frequency regions, which may be due to the codec reconstruction or resample operation applied to the original audio. While this removal does not significantly impact human perception, it has a substantial influence on CM performance. Since current deepfake speech typically produces artifacts in the high-frequency regions \cite{zhu2024slimstylelinguisticsmismatchmodel, xie2023domain, ge2022explainable}, which are crucial for CM detection.

To effectively address the high-frequency gap phenomenon, the first strategy considered is to use a low-pass filter to remove the high-frequency parts for data augmentation. This low-pass filtering strategy can reduce the disturbance caused by channel effects and codec variabilities, as has been demonstrated to be effective in ASVspoof2021\cite{wang2022low}. In practice, we use the low-pass filter from the audiomentations\footnote{https://github.com/iver56/audiomentations} for low-pass filtering. However, its reconstructed spectrogram, as shown in Figure \ref{fig:freqmask}(d), still exhibits some energy above the cutoff frequency, even when the minimum cutoff frequency is set to 0. Thus, we propose the Freqmask data augmentation method to simulate the high-frequency band gap phenomenon in evaluation set as shown in algorithm \ref{algorithm:freqmask}. 

To be specific, given an input ${x}$
x, we use the short-time Fourier transform (STFT) to convert it into a frequency domain matrix ${S}$. Then, a high-frequency cutoff is randomly selected, and values exceeding this cutoff are masked. Afterward, the inverse short-time Fourier transform (ISTFT) is applied to convert it back to the time domain as $\hat{x}$, preparing it for subsequent enhancement. In Figure \ref{fig:freqmask}, we employ the Freqmask method to enhance one training speech utterance. Specifically, the cutoff frequency is 4 kHz in Figure \ref{fig:freqmask}(b) and 7 kHz in Figure \ref{fig:freqmask}(d). From the spectrograms, it is evident that this enhancement technique closely resembles the original spectrogram shown in Figure \ref{fig:orispec}.

\begin{algorithm}
	\caption{Freqmask method for one speech utterance.}
	\label{algorithm:freqmask}
	\begin{algorithmic}[1]
		\STATE \textbf{Input:} speech utterance \( x \)
		\STATE Compute the Short-Time Fourier Transform (STFT) of \( x \):
		 \quad \( S \gets \text{STFT}(x, n\_fft, hop\_length) \)
		\STATE Compute the frequency bins:
		
		\( \text{frequencies} \gets \text{FFT\_frequencies}(sr, n\_fft) \)
		\STATE Choose a random frequency threshold \( freq \) from \(\{4000, 5000, 6000, 7000\}\)
		\STATE Find the indices of frequencies greater than \( freq \):
		\quad \( \text{high\_freq\_indices} \gets \text{Find\_Indices}(\text{frequencies}, freq) \)
		\STATE Set the magnitude of these frequencies to zero:
		\quad \( S[\text{high\_freq\_indices}, :] \gets 0 \)
		\STATE Compute the Inverse STFT to obtain the augmented speech:
		\quad \( \hat{x} \gets \text{ISTFT}(STFT, hop\_length) \)
		\STATE \textbf{Output:} augmented speech utterance \( \hat{x} \)
	\end{algorithmic}
\end{algorithm}

\begin{figure*}[t]
	\centering
	\includegraphics[width=6in]{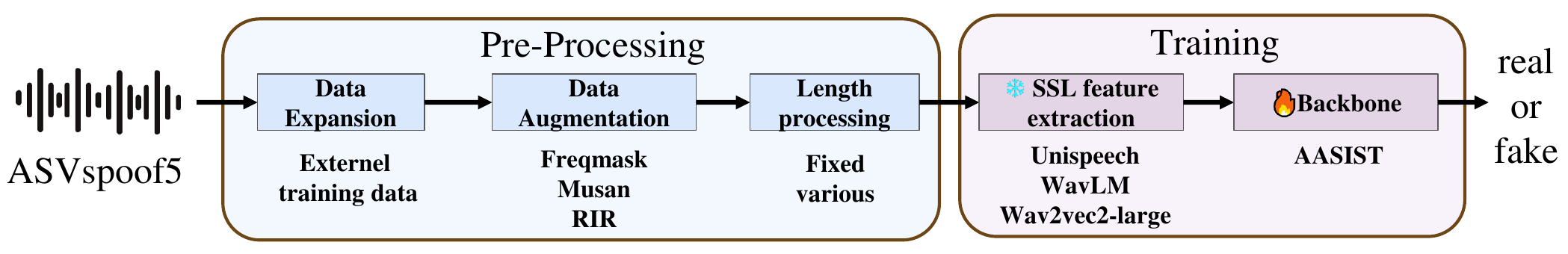}
	\caption{The duration distribution statistics of the ASVspoof5 dataset shows the audio duration on the horizontal axis and the number of occurrences in the dataset (frequency) on the vertical axis.}
	\label{fig:pipeline}
\end{figure*}

\subsubsection{Other data augmentation method}
The ASVspoof 5 dataset, utilizing MLS data as its source domain, closely resembles real-world scenarios. To better fit the real distribution, we apply DA techniques to expand the source domain data. We utilize MUSAN and RIR for noise data augmentation. Specifically, we employ the noise and music subsets of MUSAN, along with the entire RIR dataset, for augmentation. Additionally, we explored the performance of traditional DA techniques, including high-pass filtering, pitch shifting, and time stretching, using the audiomentations library on the evaluation set.

\subsection{Self-Supervised learning feature}
\label{section:ssl}
\subsubsection{SSL feature selection}
In the open condition, another key aspect is the SSL pre-trained features. The training domain for SSL features consists of a large corpus of real audio, and significant differences between real and fake speech become evident through the SSL layer. Due to the extensive size of the overall training set and limited time, we only experimented with pre-trained features by freezing specific layers rather than finetuning, as detailed below:

\textbf{WavLM \cite{chen2022wavlm}}: WavLM is built based on the HuBERT \cite{hsu2021hubert} framework, with an emphasis on both spoken content modeling and speaker identity preservation. We utilized the wavlm-base version from Hugging Face\footnote{https://huggingface.co/microsoft/wavlm-base}, which was pre-trained on 960 hours of the LibriSpeech dataset.

\textbf{Wav2vec2-large \cite{baevski2020wav2vec}}: Wav2vec2 represents a breakthrough in learning robust representations solely from speech audio. Wav2vec 2.0 masks the speech input in the latent space and solves a contrastive task defined over a quantization of the latent representations which are jointly learned. We used the wav2vec2-large version from Hugging Face\footnote{https://huggingface.co/facebook/wav2vec2-large-960h}, which was pre-trained and fine-tuned on 960 hours of 16kHz sampled speech audio from the LibriSpeech dataset.

\textbf{UniSpeech \cite{chen2022unispeech}}: UniSpeech aims to enhance the existing SSL framework for speaker representation learning. Specifically, it integrates multi-task learning and mixing strategies into the HuBERT framework, resulting in the proposed speaker-aware pre-training method, UniSpeech. We used the UniSpeech-SAT-Base version from Hugging Face\footnote{https://huggingface.co/microsoft/unispeech-sat-base}, which was pre-trained on 960 hours of LibriSpeech dataset. By extracting the hidden states of the pre-trained features, we obtained the SSL features for each audio sample, which typically have dimensions of (batch, length, dimension).

\subsubsection{Reconsider the time duration length}
\begin{figure*}[t]
	\centering
	\includegraphics[width=\linewidth]{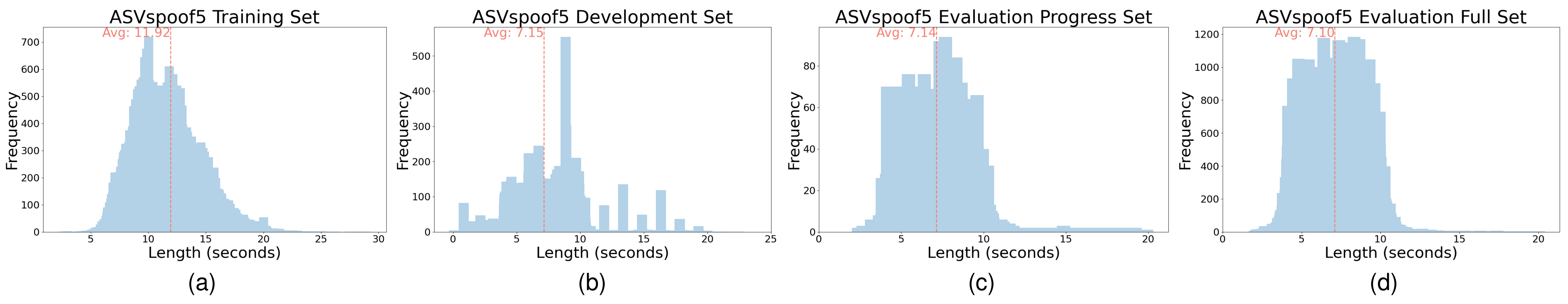}
	\vspace{-5mm}
	\caption{The duration distribution statistics of the ASVspoof5 dataset.}
	\label{fig:time}
\end{figure*}

By extracting the hidden states of the pre-trained features, we obtained the SSL features for each audio sample, which typically have dimensions of (B,L,D). B represents the batch size, L represents the audio length dimension, and D is determined by the transformer parameter of the SSL features. For the three SSL features mentioned above, the dimension of D is 768. However, the choice of audio length, L, needs to be reconsidered. As shown in Figure \ref{fig:time}, It is evident that the time-frequency distribution in this case differs from the shorter durations in 19LA, with the training set having an average duration of 11.92 seconds. Additionally, as observed from the spectrogram in Figure \ref{fig:orispec}, the silent segments in the test set are very brief. Most previous studies based on 19LA focused on the first 4 seconds for segment-level judgment, primarily benefiting from detecting the silent segments at the beginning of the audio \cite{zhang2023impact}. However, with increased audio duration in ASVspoof5, we should reconsider the selection of training and testing durations. Therefore, in our experiments, we explored the impact of duration on the CM, including segmenting or padding to the first 4 seconds, 6 seconds, 8 seconds, 10 seconds, 12 seconds, 14 seconds and 16 seconds. Additionally, we experimented with a variable-length training method, during training, each batch sequence was zero-padded to the maximum length within the batch. During inference, single-sample inference was conducted based on the actual length of each sample. 

\subsection{Backbone}
In previous SSL studies in the field of ADD, an appropriate feature selection is often much more important than the choice of the backbone network. For instance, in the 19LA experiments, using the correct SSL features, such as Wav2vec2-xls-r \cite{babu2021xls}, yields good performance even when other backbones \cite{kang2024experimental} or just FC are used \cite{lee22q_interspeech}. Therefore, in ASVspoof5 experiment, we only employed the currently the well performing backbone network in the filed of ADD, AASIST \cite{jung2022aasist}. Specifically, we used the SSL-adapted version of AASIST\cite{tak2022automatic} and adjusted the front-end dimension of the fully connected (FC) layer from the original 1024 to 768 to match our SSL dimensions. Hereto, we have comprehensively considered aspects from data expansion to the backbone. The overall pipeline is illustrated in the Figure \ref{fig:pipeline}.

\begin{table*}[!t]
	\caption{Summary of the ASVspoof5 dataset.}
	\label{tab:spoof5}
	\centering
	\setlength{\tabcolsep}{1.5mm}
	\renewcommand{\arraystretch}{1.3}
	\begin{tabular}{|c|c|c|c|c|c|c|c|}
		\hline
		Subset & \multicolumn{3}{c|}{audio length (s)} & type & real & fake & total \\
		\cline{2-4}
		& min & mean& max& & & & \\
		\hline
		Training set &2.61&11.92 &28.91 &A01-A08 &18797 &163560 &182357\\
		\hline
		Development set &0.06&7.15 &22.43 &A09-A16 &23645 &82588 &106233\\
		\hline
		Evaluation progress set &2.40&7.14 &19.91 &- &- &- &40765\\
		\hline
		Evaluation full set &2.00&7.10 &20.00 &- &- &- &680774\\
		\hline
	\end{tabular}
\end{table*}

\section{Experiments}
\subsection{Details of ASVspoof5 dataset}
Before conducting the experiments, we first analyze the ASVspoof5 dataset as shown in Table \ref{tab:spoof5}. The ASVspoof5 training set includes eight spoofing methods (A01-A08), with 18,797 real speech samples and 163,560 fake samples, totaling 182,357 audio samples, where fake samples are in the majority. The development set also includes eight methods (A09-A16), with 23,645 real samples and 82,588 fake samples, totaling 106233 fake samples. 
For the evaluation set, the organizers initially released the evaluation progress set, which contains a total of 40,765 audio samples. ASVspoof5 participants can submit results up to four times daily via the Codalab platform \footnote{https://codalab.lisn.upsaclay.fr/competitions/19380}. The Evaluation full set contains a total of 680,774 audio samples, with only one submission opportunity. The experiments and conclusion in this paper are primarily based on the results from the progress set.

In addition to the speech sample quantities, we also conducted a detailed analysis of the duration distribution of the four data subsets, as shown in the Figure \ref{fig:time}. In each subplot of Figure \ref{fig:time}, the horizontal axis represents the duration of the audio samples, while the vertical axis represents their frequency of occurrence in the dataset. From the duration distribution, we can observe that, except for the training set with an average duration of around 11 seconds, the other subsets have an average duration of around 7 seconds, which is significantly longer than the average duration of 19LA. This indicates that the duration needs to be reconsidered for CM.

\vspace{-2mm}
\subsection{Implementation details}
\vspace{-2mm}
In our experiments, the following parameters were kept constant. We used an initial learning rate of 5$\times10^{-4}$, training was performed for 20 epochs, and the learning rate was halved every 2 epochs. If a validation set is included, the model with the lowest loss on the validation set is selected as the best model; if no validation set is included, the model with the lowest loss on the training set is selected as the best model. For the loss function, due to the imbalance in the ratio of genuine to spoofed samples in the dataset, we used weighted cross-entropy, with the weights set to 10:1 for genuine and spoofed samples, respectively.

For experimental evaluation, all ASVspoof5 participants were scored using four metrics on the Codalab platform: minimum detection cost function (minDCF), actual detection cost function (actDCF), cost of log-likelihood ratio (cllr), and equal error rate (EER). Among these, minDCF is the primary ranking metric, while the other three metrics are used for reference. For all four metrics, lower values indicate better model performance. For more details on the evaluation metrics, please refer to the appendix of ASVspoof5 official evaluation plan \cite{delgado2024asvspoof}. In our experiments, we selected the primary evaluation metric of minDCF and the commonly used metric of EER in the filed of ADD to assess performance.
\begin{table}[t]
	\caption{Initial experiments for data expansion.}
	\label{tab:dataexpansion}
	\centering
	\setlength{\tabcolsep}{1.0mm} 
	\renewcommand{\arraystretch}{1.2}
	\begin{tabular}{|c|c|c|c|c|c|c|c|c|c|c|c|}
		\hline
		Dataset &DA &SSL feature &minDCF &EER
		\\
		\hline
		5train  &-  &wavlm-base-5  &0.2720 &9.50
		\\
		\hline
		5train+5dev \bf(5trndev)  &-  &wavlm-base-5 &0.1788 & 6.88
		\\
		\hline
		5trndev+19LA           &-  &wavlm-base-5 &0.1709 & 6.01
		\\
		\hline
		5trndev+MLAAD  &-  &wavlm-base-5 &0.1717 & 6.32
		\\
		\hline
		5trndev+Codecfake  &-  &wavlm-base-5 &\bf 0.1604 &\bf 5.71
		\\
		\hline
	\end{tabular}
\end{table}
\begin{table}[t]
	\caption{Results for SSL feature selection.}
	\label{tab:ssl}
	\centering
	\setlength{\tabcolsep}{1.3mm} 
	\renewcommand{\arraystretch}{1.2}
	\begin{tabular}{|c|c|c|c|c|c|c|c|c|c|c|c|}
		\hline
		Dataset &DA &SSL feature &minDCF &EER
		\\
		\hline
		5trndev  &-  &wavlm-base-last & 0.2538 & 9.28
		\\
		\hline
		5trndev&-  &wavlm-base-5  &0.1788 &6.88\\
		\hline
		5trndev&-  &wav2vec2-large-5 &0.1907 &6.63
		\\
		\hline
		5trndev  &-  &unispeech-base-5 &\bf 0.1189 &\bf 4.10
		\\
		\hline
	\end{tabular}
\end{table}

\section{Results and Discussion}
\subsection{Initial results for data expansion and feature selection}
\label{section:initial}
In the experimental setup, we initially used some CMs to determine which datasets and which SSL features to use for training. It is important to note that this experiments did not employ any DA strategies. The preliminary experiment in this section aims to validate which datasets and features can effectively enhance CM performance, paving the way for more detailed experiments later.

For the initial experiments for data expansion, we used wavlm-base-5, which is the fifth hidden layer representation of WavLM, as the feature. For SSL feature extraction, we use the first 4 seconds of audio, truncating any audio longer than 4 seconds and repeating padding to 4 seconds for audio shorter than 4 seconds. All other parameters remained consistent across the datasets. The experiment results are shown in Table \ref{tab:dataexpansion}. First, we tested using the ASVspoof5 training set (5train) as the baseline, which yielded a minDCF of 0.2720 and an EER of 9.50\%. For data expansion, before adding additional data, we first considered co-training with the validation set. Since the spoofing methods in the evaluation set are likely to be different from those in the training and validation sets (A01-A16), expanding the spoofing methods around the source domain data has been shown to be effective in previous research \cite{wang2024can}. Experiment results also proved this perspective. Co-training with 5train and ASVspoof5 development set (5dev) showed better performance on the test set compared to training with 5train alone and validating with 5dev, with minDCF decreasing by 0.0932 and EER decreasing by 2.62\%. After completing the above experiments, we decided to collectively refer to the 5train and 5dev datasets as 5trndev and to adopt a training approach without a validation set. Building on the training with 5trndev, we further explored the effects of adding 19LA, Codecfake, and MLAAD. The experimental results showed that minDCF was around 0.17 and EER was around 6\%, which did not achieve the significant improvement. On the other hand, this indicates that learning from the source domain data in 5trndev is crucial. Therefore, we decided to use the 5trndev (5train + 5dev) as the training set for the subsequent experiments.

For the initial experiments on SSL feature selection, we tested the SSL features mentioned in Section \ref{section:ssl}. We first used wavlm-base to investigate the impact of layer selection on performance. In previous experiments, the fifth layer was found to perform well in frozen feature extraction \cite{lee22q_interspeech}. Therefore, we compared the performance of the last layer and the fifth layer, as shown in the first two rows of Table \ref{tab:ssl}. The results indicate that the fifth layer significantly outperformed the last layer, with minDCF decreasing by 0.075 and EER decreasing by 2.4\%. After determining the use of the fifth layer, we experimented with wavlm-base-5, wav2vec2-large-5, and unispeech-base-5. The final results showed that unispeech-base-5 significantly outperformed the other two features, achieving the lowest minDCF of 0.1189 and the lowest EER of 4.10\%. Therefore, we chose to use unispeech-base-5 as the primary feature for subsequent experiments.

\begin{table}[t]
	\caption{Results for data augmentation. All CM used 5trndev as the training dataset and unispeech-base-5 for feature extraction.}
	\label{tab:da}
	\centering
	\setlength{\tabcolsep}{1.3mm} 
	\renewcommand{\arraystretch}{1.2}
	\begin{tabular}{|c|c|c|c|}
		\hline
		DA  &minDCF &EER
		\\
		\hline
		-                 & 0.1189 & 4.10
		\\
		\hline
		RIR               & 0.0748 & 2.86
		\\
		\hline
		RIR + MUSAN noise  & 0.0483 & 1.74
		\\
		\hline
		RIR + MUSAN & 0.0424 & 1.51
		\\
		\hline
		RIR + MUSAN + Lp & 0.0406 & 1.46
		\\
		\hline
		RIR + MUSAN + Lp + Hp & 0.0504 & 1.78
		\\
		\hline
		RIR + MUSAN + Lp + Time + Pitch & 0.0590 & 2.17
		\\
		\hline
		RIR + MUSAN + Freqmask(0.5) & 0.0388 & 1.37
		\\
		\hline
		RIR + MUSAN + Freqmask(0.3) & \bf0.0344 & \bf1.21
		\\
		\hline
	\end{tabular}
\end{table}
\begin{table}[t]
	\caption{Results for data expansion.}
	\label{tab:expansion}
	\centering
	\setlength{\tabcolsep}{0.8mm} 
	\renewcommand{\arraystretch}{1.2}
	\begin{tabular}{|c|c|c|c|c|c|c|c|c|c|c|c|c|}
		\hline
		Dataset1&Dataset2&DA1&DA2&SSL&minDCF &EER
		\\
		\hline
		5trndev  &Codecfake &RMF &- &unispeech&0.0411 & 1.43
		\\
		\hline
		5trndev  &Codecfake &RMF &RMF &unispeech&0.0437 & 1.53
		\\
		\hline
		5trndev  &19LA &RMF &- &unispeech&0.0434 & 1.50
		\\
		\hline
		5trndev  &19LA &RMF &RMF  &unispeech&0.0490 & 1.74	
		\\
		\hline
		5trndev  &MLAAD &RMF &-   &unispeech&0.0454 & 1.58
		\\
		\hline
		5trndev  &MLAAD &RMF &RMF &unispeech&0.0717 & 2.54		
		\\
		\hline		
	\end{tabular}
\end{table}
\vspace{-2mm}
\subsection{Results for data augmentation}
\vspace{-2mm}
In the experiments of this section, the training dataset is 5trndev, and the SSL feature is unispeech-base-5, which are the best settings identified in the previous subsection. The experiment results are shown in Table \ref{tab:da}. It is evident that RIR significantly enhances CM performance. With RIR, the CM achieves a minDCF of 0.0748 and an EER of 2.86\%. Next, we explored common data augmentation (DA) configurations in the field of speaker verification, combining MUSAN and RIR. This included using both the MUSAN noise subset alone and the full MUSAN set (excluding the speech subset). The experimental results show that MUSAN also provides a performance improvement for the CM, with the full MUSAN dataset further enhancing performance compared to using only the noise subset. 

After validating the effectiveness of MUSAN and RIR, we proceeded to experiment with other DA methods. The comparison results show that using Freqmask performs better than low-pass filter (LP) . Specifically, RIR + MUSAN + LP achieved a minDCF of 0.0406 and an EER of 1.46\%, while RIR + MUSAN + Freqmask(0.3) achieved the best minDCF of 0.0344 and an EER of 1.21\%. It is important to note that Freqmask(0.3) and Freqmask(0.5) represent 30\% and 50\% probabilities, respectively, for applying Freqmask to the original audio. This suggests that excessively high probabilities for applying the mask may hinder the learning of full-frequency information from the source domain. In addition to the LP, we also experimented with using high-pass filter (HP) in combination with LP, as well as time stretch and pitch shift together. However, these DA combinations did not result in further improvements in CM performance. 

\begin{table}[t]
	\caption{Results for different feature duration length.}
	\label{tab:length}
	\centering
	\setlength{\tabcolsep}{1.2mm} 
	\renewcommand{\arraystretch}{1.2}
	\begin{tabular}{|c|c|c|c|c|c|c|c|c|c|c|c|}
		\hline
		Dataset &DA &SSL feature &length&minDCF &EER
		\\
		\hline
		5trndev  &RMF  &unispeech-base-5 &4 &0.0344 & 1.21
		\\
		\hline
		5trndev  &RMF  &unispeech-base-5 &6 &0.0319 & 1.18
		\\
		\hline
		5trndev  &RMF  &unispeech-base-5 &8 &0.0284 & 0.99
		\\
		\hline
		5trndev  &RMF  &unispeech-base-5 &10 &\bf0.0254 &\bf 0.93
		\\
		\hline
		5trndev  &RMF  &unispeech-base-5 &12 &0.0291 & 1.05
		\\
		\hline		
		5trndev  &RMF  &unispeech-base-5 &14 &0.0294 & 1.02
		\\
		\hline
		5trndev  &RMF  &unispeech-base-5 &16 &0.0284 & 1.02
		\\
		\hline
		5trndev  &RMF  &unispeech-base-5 &various &0.0275 & 0.99
		\\
		\hline		
	\end{tabular}
\end{table}
\begin{table*}[t]
	\caption{Results for single CM.}
	\label{tab:cm}
	\centering
	\setlength{\tabcolsep}{1.2mm} 
	\renewcommand{\arraystretch}{1.1}
	\begin{tabular}{|c|c|c|c|c|c|c|c|c|c|c|c|}
		\hline
		CM&Dataset &DA &SSL feature &length&minDCF &EER
		\\
		\hline
		\ding{192}&5trndev  &RMF  &unispeech-base-5 &4 &0.0344 & 1.21
		\\
		\hline
		\ding{193}&5trndev  &RMF  &unispeech-base-5 &16 &0.0284 & 1.02
		\\
		\hline
		\ding{194}&5trndev  &RMF  &unispeech-base-5 &various &0.0275 & 0.99
		\\
		\hline
		
		\ding{195}&5trndev  &RMF  &unispeech-base-5 &10 &\bf0.0254 &\bf 0.93		
		\\
		\hline
		\ding{196}&5trndev  &RMF  &unispeech-base-last &10 &0.0434 & 1.50
		\\
		\hline	
		\ding{197}&5trndev  &RMF  &wavlm-base-5 &10 &0.0441 & 1.56	
		\\
		\hline
		\ding{198}&5trndev  &RMF  &wav2vec2-large &10 &0.0393 & 1.37			
		\\
		\hline
		
	\end{tabular}
\end{table*}

\begin{table}[t]
	\caption{Fusion results for different CM.}
	\label{tab:fusioncm}
	\centering
	\setlength{\tabcolsep}{1.0mm} 
	\renewcommand{\arraystretch}{1.2}
	\begin{tabular}{|c|c|c|c|c|c|c|c|c|c|c|c|}
		\hline
		Fusion type&Fusion CM&minDCF &EER
		\\
		\hline	
		Temporal &\ding{192} + \ding{193} + \ding{194} + \ding{195} &0.0193 & 0.69		
		\\
		\hline
		Feature  &\ding{195} + \ding{196} + \ding{197} + \ding{198} &0.0206 & 0.72
		\\
		\hline	
		Temporal\&Feature &\ding{192}+\ding{193}+\ding{194}+\ding{195}+\ding{196}+\ding{197}+\ding{198} &\bf0.0158 & \bf0.55
		\\
		\hline			
	\end{tabular}
\end{table}
\subsection{Results for data expansion}
Although we initially conducted preliminary data expansion experiments in section \ref{section:initial}, the selected feature was not the best, and no DA method was applied. In this section, we consider whether applying DA to the additional expanded data can further improve CM performance. The results can be seen in Table \ref{tab:expansion}, where RMF represents the best DA configuration of RIR + MUSAN + Freqmask(0.3), and Dataset2 refers to the expanded dataset. We can draw two conclusions: first, as shown in rows 1, 3, and 5 of the Table \ref{tab:expansion}, expanding extra data did not significantly improve CM performance. Second, applying DA to the external data decreased CM performance. Certainly, this only indicates that the domain information of 19LA, Codecfake, and MLAAD does not match the ASV5trndev evaluation progress set. Therefore, in subsequent experiments, we will use only the 5trndev as the training set.

\subsection{Results for different feature length}
In previous experiments, we followed the 4 seconds audio extraction scheme consistent with the 19LA series. However, this approach does not align with the duration distribution of the ASVspoof5. Therefore, we need to reevaluate the optimal duration for feature extraction based on experimental results. Specifically, we conducted experiments using fixed-length features ranging from 4 to 16 seconds. For fixed-length processing, we used the following strategy: for both training and evaluation audio, any audio longer than the specified duration was truncated, while audio shorter than the specified duration was padded by repetition to achieve the fixed length. In Table \ref{tab:length}, it is evident that as the duration of training audio increases, CM performance improves progressively. The best performance was observed with a feature duration of 10 seconds, achieving a minDCF of 0.0254 and an EER of 0.93\%. After extending the duration beyond 10 seconds, we observed no significant improvement in CM performance. This could be related to the duration distribution of the evaluation progress set, where most audio samples exceed 10 seconds as shown in Figure \ref{fig:time}(c). Additionally, it may be related to the parameters of the AASIST model. To address this, improvements should be made to handle longer duration, such as incorporating deeper ResNet convolution layers in the front-end or increasing the number of temporal graph nodes. We also explored a variable-length architecture, where max length zero-padding was applied at the batch level during training, and single-sample inference was conducted based on the actual length during evaluation. However, this strategy did not outperform the fixed 10-second scheme.
\vspace{-2mm}
\subsection{Temporal variability and multi-viewed SSL fusion for final result}
Since the fixed-length training and testing CM may disregards temporal information beyond the specified duration. For example, using 4 seconds of audio for training a CM may result in the loss of temporal information beyond 4 seconds, while using 16 seconds for training may reduce precision in evaluating the first 4 seconds. Therefore, we used CMs with different durations for score fusion, providing a multi-scaled temporal perspective for a comprehensive evaluation of audio authenticity. Specifically, we combined four different time scales as shown in the Table \ref{tab:cm}: \ding{192} 4s, \ding{195} 10s, \ding{193} 16s, and \ding{194} various durations. Each CM was assigned a fusion weight of 0.25. The fusion results, shown in the first row of Table \ref{tab:fusioncm}, achieved a minDCF of 0.0193 and an EER of 0.69\%.

In addition to temporal fusion, we also considered a multi-viewed SSL feature-level fusion. Specifically, we combined four different features: \ding{195} unispeech-base-5, \ding{196} unispeech-base-last, \ding{197} wavlm-base-5, and \ding{198} wav2vec2-large, as shown in Table \ref{tab:cm}. We selected the optimal fixed duration 10s to extract these SSL feature, and each CM was also assigned a fusion weight of 0.25. The fusion outcomes, as presented in the second row of Table \ref{tab:fusioncm}, yielded a minDCF of 0.0206 and an EER of 0.72\%.

To simultaneously integrate temporal and feature information across different CM, we fused seven types of CM, as shown in the last row of Table \ref{tab:fusioncm}. The fusion weights for optimal performance are as follows: the best performance of 10-second duration CM \ding{193} is assigned a weight of 0.3, the second best performance CM with various duration \ding{195} is assigned a weight of 0.2, and the remaining five CMs are each assigned a weight of 0.1. Our fusion system achieved a minDCF of 0.0158 and an EER of 0.55\% on the ASVspoof5 evaluation progress set. This system was also used for our final submission on the ASVspoof5 evaluation full set. However, the performance declined on the ASVspoof5 evaluation full set, with a  minDCF of 0.224 and an EER of 7.72\%.
\vspace{-5mm}
\section{Conclusion}

In this paper, we focus on the ASVspoof 5 Track 1 open condition. We initially investigate ASVspoof5 from the perspectives of data expansion, data augmentation, and SSL feature selection. Then, based on the high-frequency band gaps characteristic of ASVspoof5, we propose a DA method called Frequency Mask. Finally, by integrating multiple CM from both temporal and feature-type perspectives, we achieve a minDCF of 0.0158 and an EER of 0.55\% on the ASVspoof 5 evaluation progress set.
\vspace{-2mm}

Although our results are promising on the progress set, we observed a significant drop in performance on the ASVspoof5 evaluation full set. This discrepancy may be due to unknown deepfake methods or codec reconstruction methods present in the full set that were not appeared in the progress set. As a result, some conclusions drawn from the progress set, such as the effectiveness of additional data, optimal feature duration, and best features, may differ. In future work, we will focus on optimizing the CM performance on the evaluation full set.

\section{Acknowledgements}
This work is supported by the National Natural Science
Foundation of China (NSFC) (No.62101553).

\bibliographystyle{IEEEbib}
\bibliography{ASVspoof_BibEntries}

%

\end{document}